\documentclass[12pt,fleqn]{article}
\usepackage{t1enc}
\usepackage[latin1]{inputenc}
\usepackage{graphicx}
\usepackage[dvips]{rotating}
\begin{document}
\begin{center}
\Large {\bf PERCOLATION AND FINITE SIZE SCALING IN SEVEN DIMENSIONS }
\end{center}
\vspace{0.3cm}
\begin{center}
LOTFI ZEKRI\footnote
{Permanent address:\em U.S.T.O., D\'{e}partement de Physique, L.E.P.M., B.P.1505 El M'Naouar, Oran, Algeria. Email: lzekri@yahoo.com} 

{\em Institute for Theoretical Physics, Cologne University, D-50923 K\"oln, Germany
E-mail: zekri@thp.uni-koeln.de}

\end{center}

\vspace{0.5cm} \centerline{\bf Abstract} \bigskip

\hspace{0.5cm} Numerical investigation of critical exponents on a hypercubic lattice with
$L^{d}$ random sites with $L$ up to $33$ and $d$ up to $7$ showed that above the critical dimension the phase transitions in Ising model and percolation are not alike.

\vspace{0.5cm} 
\vfill
\vspace{0.33in}
\hspace{0.5cm}The upper critical dimension for percolation is 6, instead of 4 for the Ising model. Percolation and Ising models are special cases of the Potts model, and the five-dimensional Ising case corresponds roughly to percolation in seven dimensions \cite{01}.

\hspace{0.5cm}In the usual Ising model Cheon et al.\cite{02} based on a theory of Chen and Dohm \cite{03} calculated the spontaneous equilibrium magnetisation for $d=5$. They used finite-size scaling in the variables $L^{2}$ and $t=(T-T_c)/T_c$ according to Ref. 3, the $t - 1/L^{2}-$ plane is similar to that used in fig. 1 of Ref. 3. They found different values of the exponent for the spontaneous magnetization in different directions by approaching the critical point $1/L=0$, $t=0$. Analogous to their previous findings, we study numerically for random site percolation using the standard Hoshen-Kopelman algorithm\cite{04} the behavior of the largest cluster size and the second moment of the cluster size distribution for $d=7$ as well as $d=3$ at, below and above $p_c$. Above $p_c$ means $p-p_c =1/L^{1 / \nu}$ and below $p_c$,  $p-p_c =-1/L^{1 / \nu}$ where the connectivity length exponent $\nu$ is $0.875$  (for $d=3$) and $1/2$ (for $d=7$).

The infinite cluster normalized to the sample volume is defined as the probability $P_{\infty}$ for each site to belong that infinite cluster. This probability behaves as $P_{\infty} \propto (p-p_c)^{\beta}$. For fine size $L$,  $P_{\infty} \propto L^{D-d}$.
The second moment of the cluster size distribution $\chi$ is defined for infinite networks as $\chi \propto (p-p_c)^{-\gamma}$, and $\chi \propto (p-p_c)^{D'}$ for finite size.
Using $P_{\infty}$ for $d=7$ their exponents are summarized for different directions as follows:

\vspace{0.5cm}

\hspace{4cm} $(1/L^{2}=0, \hspace{0.2cm} p>p_c )$ ; $\beta \approx 1$ , $\gamma=1$ \hspace{4cm} $(1a)$

\vspace{0.5cm} \hspace{4cm} $(1/L^{2}>0, \hspace{0.2cm}  p=p_c )$ ; $D=4.2$ , $D'=1.8$ \hspace{3cm} $(1b)$

\vspace{0.5cm} \hspace{4cm} $(p-p_c=1/L^{2} )$ ; $D=5$ , $D'=2$ \hspace{4cm} $(1c)$

\vspace{0.5cm} \hspace{4cm} $(p-p_c=-1/L^{2} )$; $D=4.2$ , $D'=2$ \hspace{3.5cm} $(1d)$

\vspace{0.5cm}

and using the relations\cite{04}:

\vspace{0.5cm} 
\hspace{4cm} $ D=d-\beta/\nu $  \hspace{7.6cm} $(2)$

\vspace{0.5cm} \hspace{4cm} $ D'=\gamma/\nu$   \hspace{8.2cm} $(3)$

\vspace{0.5cm}  we obtain the same exponents in the directions $a$ and $c$. For the Ising model tested by Ref. 2, the powers for the directions $b$, $c$ and $d$ are respectively $1/L^{5/4}$, $1/L$ and $1/L^{3/2}$.

\hspace{0.5cm} Now let us describe the behavior of these exponents in detail. We plot in figure 1 the maximum cluster exponent $D$ for $d=7$ along directions $b$, $c$ and $d$. This exponent seems to be close to 4.2 at $p_c$ and also below this critical point with very large fluctuations, whereas above this point the exponent seems to be very close to 5 with very small fluctuations. The fluctuations can be explained by the existence of a very large number of small clusters below $p_c$ and the critical behavior at this phase transition, while above $p_c$ the largest cluster size is very large so that it fluctuates only slightly.
\begin{figure}[h!]
\centering
\includegraphics[width=9cm, angle=-90]{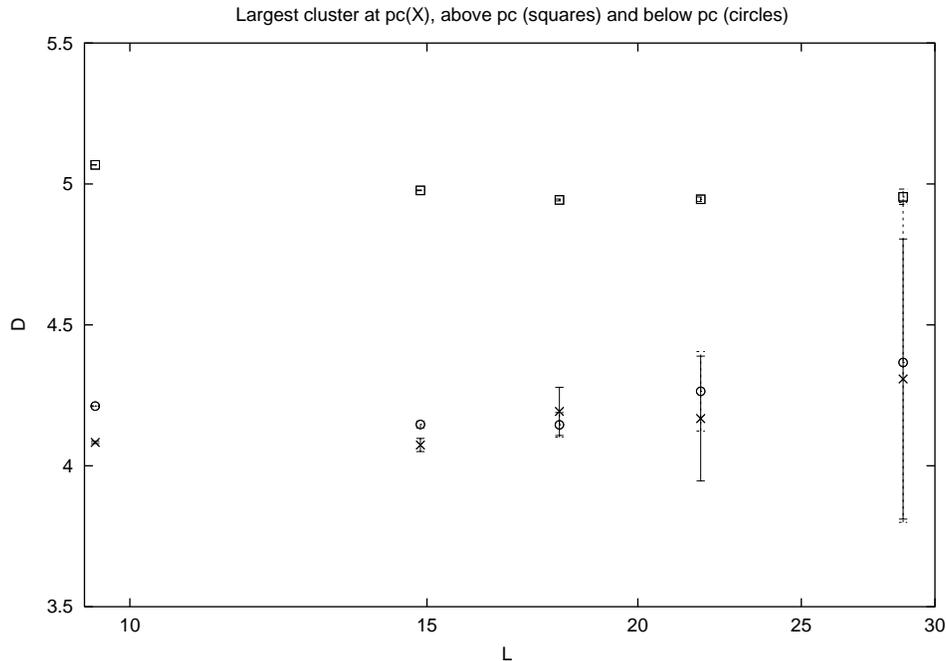}
\parbox{11cm}{\caption {{\footnotesize Semi-log plot of the maximum cluster scaling exponent $D$ for $d=7$. We use sizes from 7 until 33, $64 \times 10^{5}$ samples for $L=7$, 64000 samples for $L=13$, 12800 samples for $L=17$, 6400 samples for $L=19$, 60 samples for $L=25$ and 10 samples for $L=33$.}}}  
\label{fig1}
\end{figure}
\begin{figure}[t!]
\centering
\includegraphics[width=9cm,angle=-90]{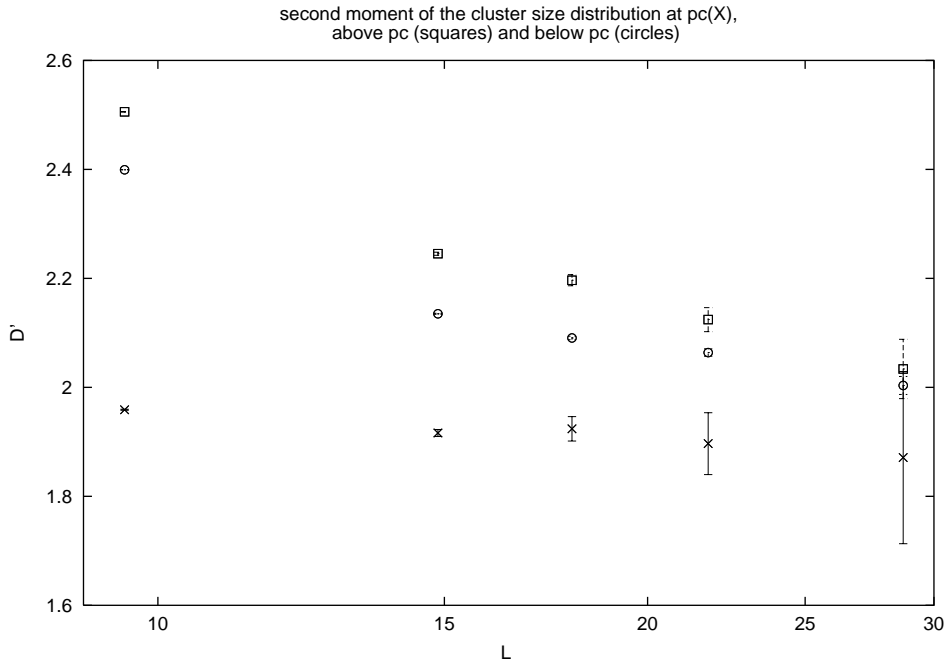}
\parbox{12cm}{\caption{{\footnotesize Semi-log plot of the exponent $D'$ for $d=7$ and the same sizes and statistic cited in figure 1}}}
\label{fig2}
\end{figure}

We show in figure 2 the exponent $D'$ for the second moment of the cluster size distribution (ignoring the largest cluster). At both sides of $p_c$ this exponent is close to 2, while it seems to be close to 1.8 at $p_c$ with large fluctuations for the same reasons as in figure 1. We cannot compare our data with \cite{05} since there all spanning clusters, not one largest cluster, is studied. 

\hspace{0.5cm}In order to check the theoretical prediction of $D$ and $D'$ for dimension 3 using our Monte Carlo simulation we calculated with less computational efforts these exponents for $L$ up to 401 and 500 samples. They seem to not depend on the direction and fluctuate strongly around 2.5 and 2 not far from the theoretical values 2.53 and 2.06. 
\begin{figure}[h!]
\centering
\includegraphics[width=9cm,angle=-90]{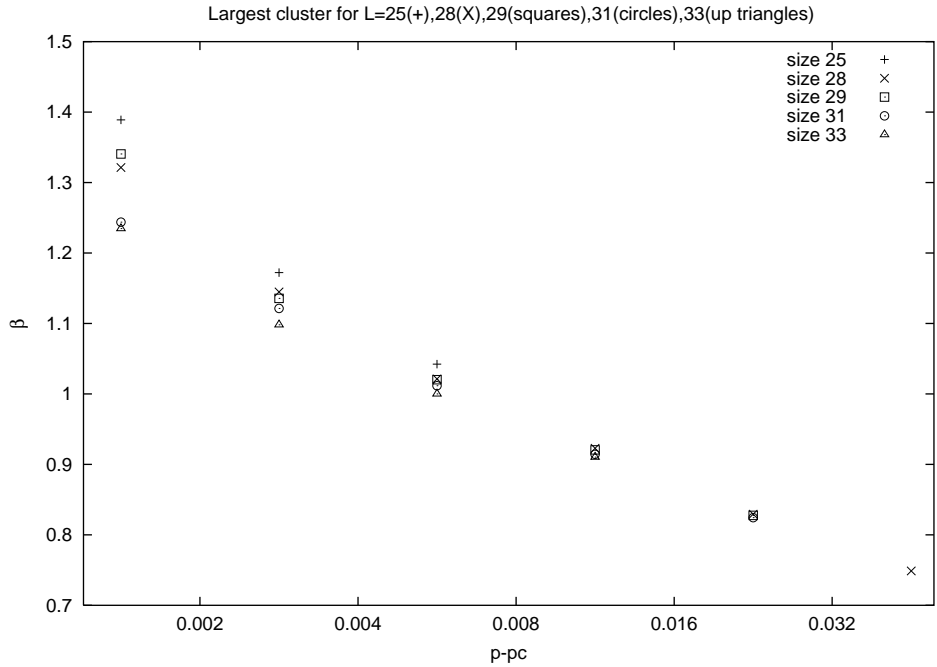}
\parbox{12cm}{\caption{{\footnotesize The largest cluster exponent $\beta$ of the infinite cluster versus $p-p_c$ for $d=7$. $L=33$ is the largest size we can simulate, 25 samples
for $L=25$ and one sample for the other sizes.}}}
\label{fig3}
\end{figure}
\begin{figure}[h!]
\centering
\includegraphics[width=9cm,angle=-90]{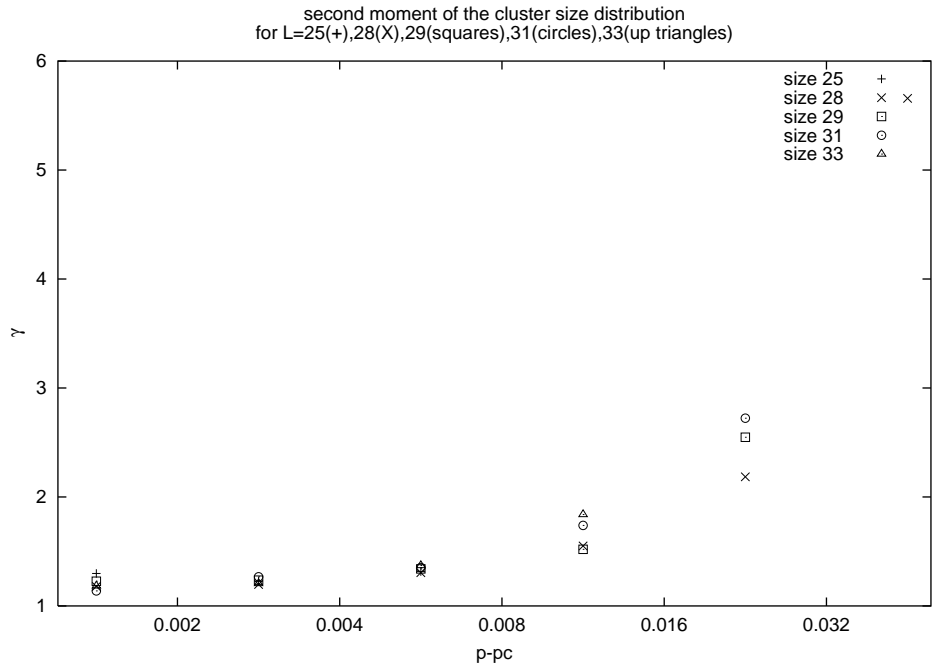}
\parbox{12cm}{\caption{{\footnotesize The second moment of the cluster size distribution exponent $\gamma$ versus $p-p_c$ for $d=7$ with the same sizes and statistic of figure 3}}}
\label{fig4}
\end{figure}

\hspace{0.5cm} Finally, we check the theoretical values of $\beta$ and $\gamma$ along the direction $a$. We show in figures 3 and 4 the exponent $\beta$ for the largest cluster and the exponent $\gamma$ for the second moment of the cluster size distribution for $p_c$ approached from above for $d=7$ at fixed large $L$. In figure 3 the exponents cannot be distinghuished above $p-p_c=0.008$ for all sizes used. But close to $p_c$ they differ for $\beta$ by several percent from each other and by about 20 percent from the theoretical value 1. In contrast, in figure 4, the exponent $\gamma$ is in a good agreement with the theoretical value 1 close to $p_c$.

\vspace{1cm}
Conclusion:

\hspace{0.5cm} We showed in this work that percolation and Ising models behave differently above the critical dimensions in the size range investigated. 
Because of finite-size effects, the exponents should not be expected to agree perfectly  with theory.

\vspace{0.5cm}
{\bfseries Acknowledgments}

\vspace{0.2cm}
I wish to thank D. Stauffer for helpful
discussions, the Julich supercomputer center for CPU time provided in their Cray T3E and IBM Jump and DAAD for financial support during the progress of this work.

\end{document}